\documentclass[aps,prd,notitlepage,reprint,twocolumn,showpacs,superscriptaddress,longbibliography,nofootinbib,floatfix]{revtex4-1}

\usepackage{amsmath,amssymb,amsfonts}
\usepackage{graphicx}
\usepackage{verbatim}
\usepackage{pifont}

\usepackage{hyperref}
\usepackage{slashed}
\usepackage{verbatim}

\usepackage[usenames]{color}
\definecolor{purple}{rgb}{0.5,0,0.8}

\newcommand{\be}{\begin{equation}}
\newcommand{\ee}{\end{equation}}
\newcommand{\bi}{\begin{itemize}}
\newcommand{\ei}{\end{itemize}}
\newcommand{\bea}{\begin{eqnarray}}
\newcommand{\eea}{\end{eqnarray}}
 
\newcommand{\bra}[1]{\langle\,#1\,|}          
\newcommand{\ket}[1]{|\,#1\,\rangle}          
\newcommand{\ud}{\mathrm{d}}

\newcommand{\LCm}{{\scriptscriptstyle -}} 
\newcommand{\LCp}{{\scriptscriptstyle +}}

\newcommand{\LCperp}{{\scriptscriptstyle \perp}}

\newcommand{\LCpara}{{\scriptscriptstyle \parallel}}

\usepackage[T1]{fontenc} 

\usepackage{bm}

\begin{document}

\title{Thermally versus dynamically assisted Schwinger pair production}

\author{Greger Torgrimsson}
\email{greger.torgrimsson@uni-jena.de}
\affiliation{Theoretisch-Physikalisches Institut, Abbe Center of Photonics,
	Friedrich-Schiller-Universit\"at Jena, Max-Wien-Platz 1, D-07743 Jena, Germany}
\affiliation{Helmholtz Institute Jena, Fr\"obelstieg 3, D-07743 Jena, Germany}

\begin{abstract}

We study electron-positron pair production by the combination of a strong, constant electric field and a thermal background. We show that this process is similar to dynamically assisted Schwinger pair production, where the strong field is instead assisted by another coherent field, which is weaker but faster. We treat the interaction with the photons from the thermal background perturbatively, while the interaction with the electric field is nonperturbative (i.e. a Furry picture expansion in $\alpha$). 
At $\mathcal{O}(\alpha^2)$ we have ordinary perturbative Breit-Wheeler pair production assisted nonperturbatively by the electric field. Already at this order we recover the same exponential part of the probability as previous studies, which did not expand in $\alpha$.
This means that we do not have to consider higher orders, so our approach allows us to calculate the pre-exponential part of the probability, which has not been obtained before in this regime. Although the prefactor is in general subdominant compared to the exponential part, in this case it can be important because it scales as $\alpha^2\ll1$ and is therefore much smaller than the prefactor at $\mathcal{O}(\alpha^0)$ (pure Schwinger pair production). 
We show that, because of the exponential enhancement, $\mathcal{O}(\alpha^2)$ still gives the dominant contribution for temperatures above a certain threshold, but, because of the small prefactor, the threshold is higher than what the exponential alone would suggest.

\end{abstract}
\maketitle

\section{Introduction}

Pure Schwinger pair production~\cite{Sauter:1931zz,Heisenberg:1935qt,Schwinger:1951nm} by a constant electric field alone is unlikely to be observed any time soon, 
but there are non-spontaneous processes which have similar nonperturbative features and could occur at much lower intensities. One example is trident pair production $e^\LCm\to2e^\LCm+e^\LCp$~\cite{Baier,Ritus:1972nf,Bamber:1999zt,Hu:2010ye,Ilderton:2010wr,King:2013osa,Dinu:2017uoj,King:2018ibi,Mackenroth:2018smh}. This requires much lower intensities because in the rest frame of a high-energy electron the field strength is much higher, and in the semiclassical regime this process has a similar nonperturbative exponential behavior as the Schwinger mechanism~\cite{Baier,Ritus:1972nf,Bamber:1999zt,Dinu:2017uoj}. If one prefers to keep the initial state massless, one can instead significantly enhance the probability by sending a high-energy photon through the electric field~\cite{Dunne:2009gi}. Another way to enhance the probability is to add a second coherent field, which is weaker but faster~\cite{Schutzhold:2008pz}. The latter is referred to as dynamically assisted Schwinger pair production and has been studied in many papers in the last decade, see e.g.~\cite{Schutzhold:2008pz,Orthaber:2011cm,Otto:2014ssa,Linder:2015vta,Schneider:2016vrl,Torgrimsson:2017pzs,Torgrimsson:2017cyb,Aleksandrov:2018uqb,Torgrimsson:2018xdf}.

Another interesting question is how Schwinger pair production (and the effective action) is affected by a nonzero temperature, see e.g.~\cite{Dittrich:1979ux,Thermal1980,Kim:2008em,Cox:1984vf,Loewe:1991mn,Elmfors:1994fw,Hallin:1994ad,Ganguly:1995mi,Shovkovy:1998xw,Gies:1998vt,Gies:1999vb,Gavrilov:2006jb,Kim:2007ra,Gavrilov:2007hq,Kim:2008em,Monin:2009aj,Kim:2010qq,King:2012kd,Fukushima:2014sia,Brown:2015kgj,Medina:2015qzc,Gould:2017fve,Korwar:2018euc,Draper:2018lyw,Gould:2018ovk,Sheng:2018jwf,Gould:2018efv}. It is fair to say that thermal pair production is a somewhat controversial topic with many papers that disagree with each other. In this paper we are interested in regimes where the thermal background leads to an exponential increase in the probability as in~\cite{Brown:2015kgj,Gould:2017fve}. In this paper we only consider thermal photons. One might expect that effects from thermal fermions are suppressed at low temperatures, or one could imagine somehow filtering out the fermions~\cite{Gould:2018efv}, as we are only interested in the thermal distribution right before the field is applied. In any case, this is enough to study the exponential enhancement in~\cite{Brown:2015kgj,Gould:2017fve}, which we will show is very similar to dynamical assistance, by comparing with the approach in~\cite{Torgrimsson:2017pzs,Torgrimsson:2017cyb,Torgrimsson:2018xdf}. 

Since Schwinger pair production is nonperturbative in the field strength and since the additional weak field in dynamical assistance is also coherent, it might not have been obvious how the probability in dynamical assistance depends on the weak field. Even if not a nonperturbative dependence, one might have thought that one would in general have to calculate too many orders for an expansion in the field strength of the weak field to be useful. However, we have showed that it is in many cases useful to study dynamical assistance by such a power series expansion~\cite{Torgrimsson:2017pzs,Torgrimsson:2017cyb,Torgrimsson:2018xdf}. For the purpose of this paper, this is best illustrated with a weak field in the shape of a Sauter pulse. So, consider an electric field given by $E_z(t)=E(f_0(t)+\varepsilon f(t))$, where $E\ll1$\footnote{Throughout this paper we use units with $c=\hbar=m_e=k_{\rm B}=1$ and we rescale the field strength $eE\to E$, where $m_e$ and $e$ are the electron mass and charge.} and $f_0=1$ are the field strength and field shape of the strong and approximately constant field, and $\varepsilon\ll1$ is the relative field strength of the weaker field. For a Sauter pulse we have $f(t)=1/\cosh^2(\omega t)$. By treating both the strong and the weak field together with worldline instanton or WKB methods one finds a probability with the following exponential part~\cite{Schutzhold:2008pz}
\be\label{SauterExponent}
P\sim\exp\left\{-\frac{2}{E}\left(\frac{\sqrt{\gamma_*^2-1}}{\gamma_*^2}+\text{arcsin}\frac{1}{\gamma_*}\right)\right\} \;,
\ee   
where $\gamma_*=\gamma/\gamma_{\rm crit}$, $\gamma=\omega/E$ is the Keldysh parameter and $\gamma_{\rm crit}=\pi/2$. For $\gamma_*>1$ this gives an exponential enhancement of the probability. Note that in those approaches this exponent is obtained from an expression that initially includes the field strength of the weak field, but the final result~\eqref{SauterExponent} is independent of $\varepsilon$. In~\cite{Torgrimsson:2017pzs} we showed that the exponent in~\eqref{SauterExponent} can also be obtained by treating the weak field perturbatively. In fact, we find this exponent already at the first order, i.e. from the absorption of a single photon from the weak field. We also showed that all the higher orders have the same exponential. Since the higher orders have higher powers of $\varepsilon\ll1$ this means that the first order gives the dominant contribution for this field. The reason that this happens for a Sauter pulse can be understood from its Fourier transform, which at large Fourier frequencies scales as
\be
f(\omega_1)\sim e^{-\frac{|\omega_1|}{\omega_*}}
\ee    
for $|\omega_1|\gg\omega_*$ where $\omega_*=2\omega/\pi$. This exponential decay is a slow decay, i.e. the Fourier transform is wide, which means that the suppression of the Fourier transform at large Fourier frequencies is less important than the suppression due to higher powers of the perturbative expansion parameter, so the first order gives the dominant contribution. The fact that we do not have to calculate higher orders of course makes the calculations simpler and we have found very good agreement with the exact numerical result~\cite{Torgrimsson:2017pzs}. From an experimental point of view it is important to notice that even if the characteristic frequency is well below the electron mass, $\omega\ll1$, the Fourier frequencies that give the dominant contribution are on the order of the electron mass,
\be
\omega_1^{\rm dom}=2\sqrt{1-\frac{1}{\gamma_*^2}}=\mathcal{O}(1) \;.
\ee 

\begin{figure}
\includegraphics[width=\linewidth]{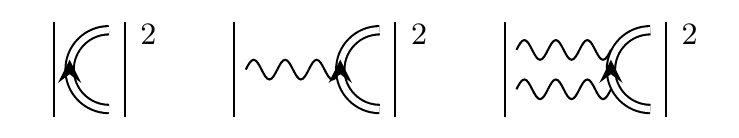}
\caption{The first three terms in the Furry picture expansion. Double lines represent fermions dressed by the electric background field. Wiggly lines are photons from the thermal background.}
\label{BWampFig}
\end{figure}

While a Sauter pulse might not be the most realistic field shape, it is, as noted in~\cite{Torgrimsson:2018xdf}, an example of a field which leads to the closest connection to thermally assisted pair production. In~\cite{Brown:2015kgj,Gould:2017fve} the exponential part of the probability of pair production by a constant electric field at temperature $T$ was obtained, and the result has exactly the same functional form as in~\eqref{SauterExponent} for dynamical assistance, but with $\gamma_*=2T/E$. In this paper we will show that this close similarity means that we can study thermal assistance with essentially the same methods as the ones we used in~\cite{Torgrimsson:2017pzs,Torgrimsson:2017cyb,Torgrimsson:2018xdf} for dynamical assistance. Here it is the usual fine-structure constant $\alpha=e^2/(4\pi)$ that is the perturbative expansion parameter, so this is basically a Furry-picture expansion where the electric field is taken into account nonperturbatively. The first three terms are shown in Fig.~\ref{BWampFig}. 
The zeroth order gives the usual Schwinger mechanism without any enhancement~\cite{Elmfors:1994fw,Gies:1998vt,Gies:1999vb} \footnote{However, if one takes into account thermal fermions then the Pauli principle leads to a reduction of $\mathcal{O}(\alpha^0)$~\cite{Thermal1980,Gavrilov:2006jb,Kim:2008em,Gavrilov:2007hq,Fukushima:2014sia,Sheng:2018jwf}.}, and the higher orders lead to exponential enhancement due to the absorption of thermal photons. In comparison with previous studies of the effective action, see e.g.~\cite{Gies:1999vb}, note that the $n$-th order corresponds to $(n+1)$-loop diagrams. 
Note also that the thermal background describes the content of photons in the initial state, before the electric field is switched on.

The rest of this paper is organized as follows. In Sec.~\ref{FirstOrderSection},~\ref{SecondOrderSection} and~\ref{HigherOrdersSection} we calculate $\mathcal{O}(\alpha)$, $\mathcal{O}(\alpha^2)$ and higher orders, respectively. For $\mathcal{O}(\alpha)$ and $\mathcal{O}(\alpha^2)$ we calculate both the exponential and the prefactor part of the probability. In Sec.~\ref{HigherOrdersSection} we show that the higher orders have the same exponential as $\mathcal{O}(\alpha^2)$, which means that we do not have to calculate the prefactor at higher orders.

\section{First order}\label{FirstOrderSection}

At first order we have pair production assisted by a single thermal photon, illustrated by the second diagram in Fig.~\ref{BWampFig}. In dynamical assistance the exponent in~\eqref{SauterExponent} is generated by off-shell photons with zero spatial momentum. So, it seems already clear that a single on-shell thermal photon will not give~\eqref{SauterExponent}. It is easy to check that it indeed gives something different. We start with the result in~\cite{Dunne:2009gi} for pair production by a single on-shell photon in a constant electric field, which is given by
\be
P_\omega\sim{\dots}\exp\left\{-\frac{2}{E}\left((1+p^2)\text{arctan}\frac{1}{p}-p\right)\right\} \;,
\ee 
where $p=|\sin\theta|\omega/2$, $\omega$ is the frequency of the photon, $\theta$ is the angle between the field and the direction in which the photon travels, and the ellipses denote the prefactor which can be found in~\cite{Dunne:2009gi}. As in~\cite{King:2012kd}, the probability of pair production by a single thermal photon is given by
\be\label{P1thermalSum}
P_1=\sum_{\rm pol.}\int\frac{\ud^3{\bf k}}{(2\pi)^3}\frac{1}{e^{\omega/T}-1}P_\omega \;,
\ee
where ${\bf k}$ is the photon momentum and $1/(e^{\omega/T}-1)$ is the photon density. We need high frequencies for significant enhancement and we are interested in not too high temperature 
$T\ll1$, so we can approximate $1/(e^{\omega/T}-1)\approx e^{-\omega/T}$ and perform the momentum integral with the saddle-point method. The exponent is maximized at $\theta=\pi/2$ and a frequency that is determined by a transcendental saddle-point equation (cf.~Eq.~(7) in~\cite{Torgrimsson:2017cyb}), 
\be
1-p\text{ arctan}\frac{1}{p}=\frac{1}{\gamma} \;,
\ee
which we solve numerically and substitute in
\be\label{P1p}
P_1=V_4\frac{\alpha(\gamma E)^2}{8\pi}\frac{p(1+3p^2)}{\sqrt{\gamma-1}(\gamma-1-p^2)}e^{-\frac{2}{Ep}\left(1-\frac{1-p^2}{\gamma}\right)} \;.
\ee
\begin{figure}
\includegraphics[width=\linewidth]{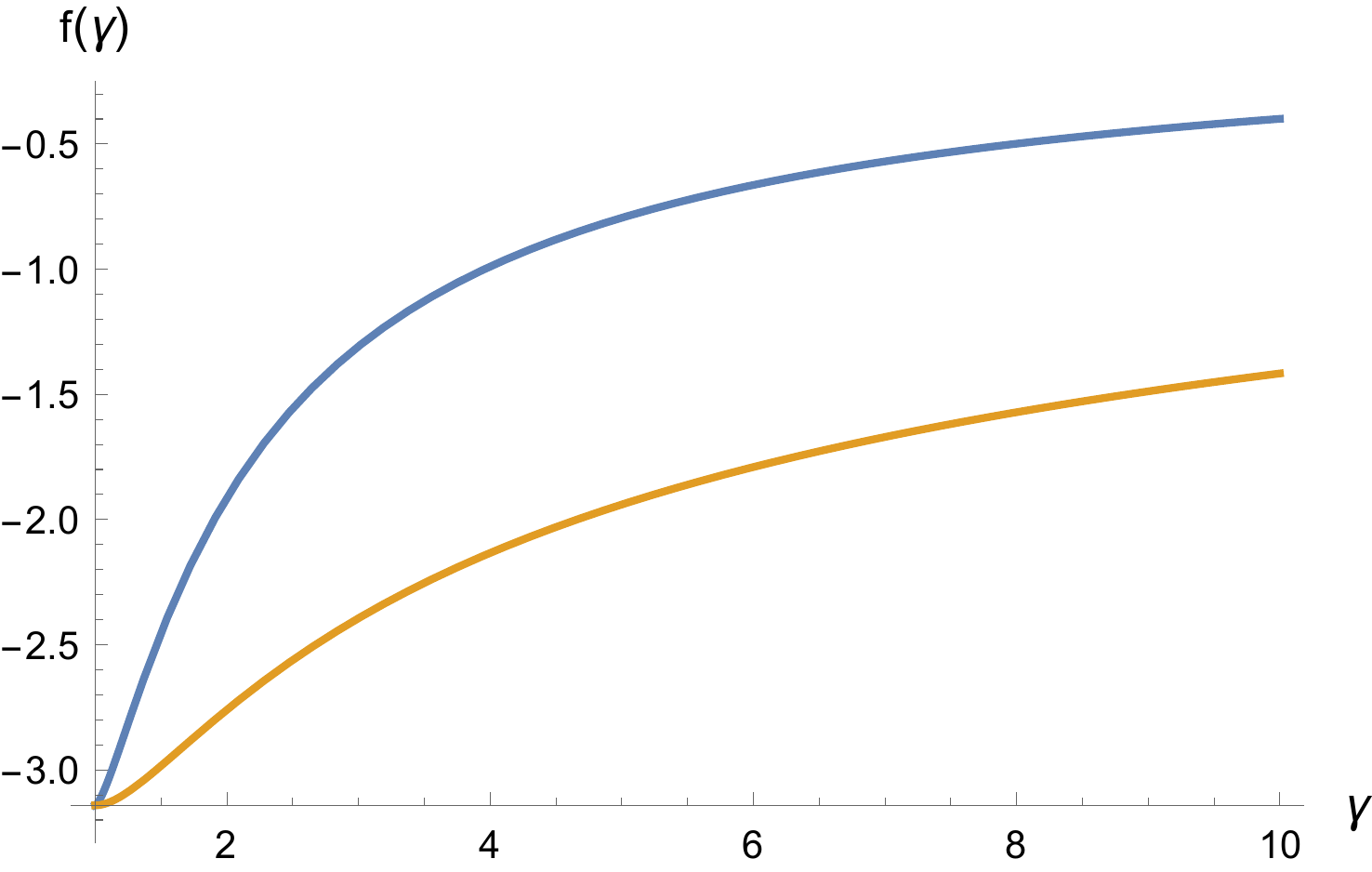}
\caption{Pair production probability $P\sim e^{f(\gamma)/E}$, where $\gamma=2T/E$. The orange curve shows the exponent in~\eqref{P1p} and the blue curve shows~\eqref{SauterExponent}.}
\label{firstOrderPlot}
\end{figure}
As shown in Fig.~\ref{firstOrderPlot} this leads to a smaller exponential compared to~\eqref{SauterExponent}, so its contribution to the probability is much smaller for $E\ll1$. In the limit $\gamma\gg1$ we find
\be\label{P1high}
P_1\to V_4\frac{\sqrt{3}\alpha T^2}{4\pi}\exp\left\{-\frac{8}{E\sqrt{3\gamma}}\right\} \;,
\ee 
which vanishes in the limit $E\to0$. (The exponential part of~\eqref{P1high} has the same form as Eq.~(9) in~\cite{Torgrimsson:2017cyb}, but with $\gamma_{\rm crit}=1$.) Note that~\eqref{P1high} is always nonperturbative in $E$, in contrast to the $\gamma\gg1$ limit of~\eqref{SauterExponent}, which scales as $e^{-2/T}$. Thus, the exponential scaling of $P_1$ is significantly different from~\eqref{SauterExponent}.

\section{Second order}\label{SecondOrderSection}

At second order we have ordinary perturbative Breit-Wheeler pair production by two thermal photons assisted nonperturbatively by the electric field, illustrated by the third diagram in Fig.~\ref{BWampFig}. Perturbative Breit-Wheeler at zero field was studied in~\cite{Gould:2018efv}\footnote{They also studied the thermal-field combination, but with a different method.} and the exponential part in the zero-temperature case was studied in~\cite{Satunin:2018rdw}. As far as we are aware, this is the first time that the combination of both is studied.
Here two photons are absorbed from the thermal background. While both are on-shell their combined momentum can be off-shell with zero spatial momentum, and this gives the dominant contribution. In~\cite{Torgrimsson:2018xdf} we showed how to calculate dynamical assistance at second order and higher. Here we can use essentially the same methods.
This perturbative approach may in fact be even more useful here, because, while one in dynamical assistance can obtain the exact ($\mathcal{O}(\alpha^0)$) result by numerically solving the Dirac equation in both the strong and the weak field, see e.g.~\cite{Schneider:2016vrl,Aleksandrov:2018uqb}, there is no corresponding exact numerical approach for thermal assistance.


\subsection{Derivation}

The probability is given by (cf.~\cite{Gould:2018efv,Weaver76} for the corresponding thermal sum in the purely perturbative case) 
\be\label{P2thermalSum}
\begin{split}
P_2=\frac{1}{2}&\sum_{\rm pol.}\int\frac{\ud^3{\bf k}_1}{(2\pi)^3}\frac{\ud^3{\bf k}_2}{(2\pi)^3}\frac{1}{e^{\omega_1/T}-1}\frac{1}{e^{\omega_2/T}-1} \\
&\sum_{\rm spin}\int\frac{\ud^3{\bf p}}{(2\pi)^3}\frac{\ud^3{\bf p}'}{(2\pi)^3}|M_2|^2 \;,
\end{split}
\ee   
where the factor of $1/2$ prevents double counting of identical particles, ${\bf k}_1$ and ${\bf k}_2$ are the momenta of the two photons, and ${\bf p}$ and ${\bf p}'$ are the momenta of the produced electron and positron, respectively. The amplitude can be written as
\be
\begin{split}
M_2=&\frac{(-ie)^2}{\sqrt{2\omega_1 2\omega_2}}\int\ud^4x_1\ud^4x_2\bar{u}_{s,{\bf p}}(t)e^{ip_jx_1^j}\slashed{\epsilon}_1e^{-ik_1x_1} \\
&iG(x_1,x_2)\slashed{\epsilon}_2e^{-ik_2x_2}v_{s',{\bf p}'}(t')e^{ip'_jx_2^j}+(1\leftrightarrow2)
\end{split} \;,
\ee 
where $\epsilon_\mu({\bf k})$ denotes a polarization vector, $(1\leftrightarrow2)$ is obtained from the first term by swapping place of the two photons, and the electric field enters via the electron and positron spinors, $u$ and $v$, and the propagator $G$. The exact propagator is given by~\cite{Schwinger:1951nm,Fradkin:1991zq,Dittrich:2000zu}
\be\label{PropagatorInConstantE}
\begin{split}
G(x,x')=-&e^{-\frac{iE}{2}(z-z')(t+t')}\int\frac{\ud^4q}{(2\pi)^4}e^{-iq(x-x')} \int_0^\infty
\ud s \\
&\exp\left\{-sm_\LCperp^2+(q_0^2-q_3^2)\frac{\tan(Es)}{E}\right\} \\
&\Big[\slashed{q}+m+i(\gamma^0q_3+\gamma^3q_0)\tan(Es)\Big] \\
&\Big[1-i\gamma^0\gamma^3\tan(Es)\Big] \;,
\end{split}
\ee
where $m_\LCperp=\sqrt{1+q_\LCperp^2}$ and $q_\LCperp=\{q_1,q_2\}$. 
The spinors can of course also be obtained exactly in a constant electric field, but here we only need the corresponding WKB approximations, which are given by (see e.g.~\cite{Hebenstreit:2011pm,Hebenstreit:2010vz,Torgrimsson:2017pzs}) 
\be\label{UandV}
\begin{split}
	U_r(t,{\bf q})&=(\gamma^0\pi_0+\gamma^i\pi_i+1)G^+(t,{\bf q})R_r \\
	V_r(t,-{\bf q})&=(-\gamma^0\pi_0+\gamma^i\pi_i+1)G^-(t,{\bf q})R_r \;,
\end{split}
\ee
where $\pi_\LCperp=q_\LCperp$, $\pi_3(t)=q_3-A(t)$, $\pi_0=\sqrt{m_\LCperp^2+\pi_3^2(t)}$, $r=1,2$, $\gamma^0\gamma^3 R_s=R_s$ and
\be\label{Gpmdef}
G^\pm(t,{\bf q})=[2\pi_0(\pi_0\pm\pi_3)]^{-\frac{1}{2}}\exp\bigg[\mp i\int_0^t\!\ud t'\,\pi_0(t')\bigg] \;,
\ee
where the lower integration limit is arbitrary, and for a constant field we have
\be\label{intpi0tophi}
\int_0^t\pi_0=-\frac{m_\LCperp^2}{2E}\left(\phi\left[\frac{p_3-Et}{m_\LCperp}\right]-\phi\left[\frac{p_3}{m_\LCperp}\right]\right) \;,
\ee
where
\be
\phi(u)=u\sqrt{1+u^2}+\text{arcsinh }u \;.
\ee  

We start by performing the trivial spatial integrals. These give the overall momentum conservation $(2\pi)^2\delta^3({\bf p}+{\bf p}'-{\bf k}_1-{\bf k}_2)$ and another delta function which we use to perform the ${\bf q}$ integrals in the propagator. The square of the overall momentum delta function gives a spatial volume factor $V_3=(2\pi)^3\delta^3(0)$ and a delta function which we use to perform the ${\bf p}'$ integrals. The $s$ integral in the propagator receives the dominant contribution at $s=\mathcal{O}(E^0)$, so apart from the $e^{-\frac{iE}{2}(z-z')(t+t')}$ factor the propagator reduces to the field-free one. We can again approximate $1/(e^{\omega/T}-1)\approx e^{-\omega/T}$. At the amplitude level we now have an exponential given by
\be
e^{-\frac{\omega_1+\omega_2}{2T}+i\int_0^{t_1}\!\pi_0({\bf p})-i\omega_1 t_1-iq_0(t_1-t_2)-i\omega_2 t_2+i\int_0^{t_2}\!\pi_0(-{\bf p}')} \;.
\ee  
We change variables $t_1\to(m_\LCperp\tau_1+p_3)/E$ and $t_2\to(m'_\LCperp\tau_2-p'_3)/E$, where $m'_\LCperp=\sqrt{1+{p'}_\LCperp^2}$, and to ${\bf\Sigma}=({\bf k}_2+{\bf k}_1)/2$ and ${\bf\Delta}={\bf k}_2-{\bf k}_1$. We perform the $\tau_i$, $q_0$, $p_\LCperp$, ${\bf K}$ and $|{\bf\Delta}|$ integrals with the saddle-point method. We have a saddle point at $\tau_i=i/\gamma$, $q_0=0$, $p_\LCperp=0$, ${\bf\Sigma}=0$ and $|{\bf\Delta}|=2\sqrt{1-\frac{1}{\gamma^2}}$, where $\gamma=2T/E$. 
This means that we are considering the region close to the point where the pair is produced without a heavier effective mass ($m_\LCperp=m'_\LCperp=1$) by two photons colliding head on (${\bf k}_2=-{\bf k}_1$), and because the photons are assisted by the field, they have energies below the mass gap ($\omega_1=\omega_1=\sqrt{1-\frac{1}{\gamma^2}}<1$), but still close to it ($\omega_1=\omega_2\sim1$). The perturbation around this point contributes to the prefactor. 
To calculate the spinor part of the prefactor we have used an explicit basis for $\gamma^\mu$ and $R_r$ as in~\cite{Torgrimsson:2017pzs,Torgrimsson:2018xdf}. 
The summation over photon polarization can be done either by choosing explicit vectors $\epsilon_\mu$ or as in the standard free-field case. In spherical coordinates for ${\bf\Delta}$ we find that the integral over the angle between ${\bf\Delta}$ and the electric field is elementary and the other angular integral is trivial. The integrand does not depend on $p_3$ so, as is well known, it then gives a temporal volume factor $\int\ud p_3=EV_0$. 

As mentioned, for thermal assistance there are no exact numerical methods to compare with. However, in~\cite{Torgrimsson:2017pzs,Torgrimsson:2017cyb,Torgrimsson:2018xdf} we have showed that the corresponding (e.g. saddle-point) approximations for dynamical assistance agree well with the exact numerical result in the regimes that we are interested in here, and, because of the close similarity, those comparisons also give us a sense of the accuracy of the approximations presented here.

Another way to derive the same result is to use unitarity to obtain the pair production probability from loops with four photon vertices.
\begin{figure}
\includegraphics[width=\linewidth]{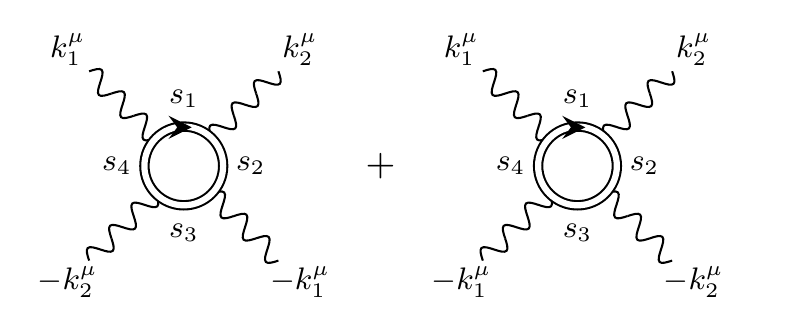}
\caption{The real part of these loops gives via unitarity the pair production probability.}
\label{loopsFig}
\end{figure}
Let $\mathcal{M}$ be the amplitude for two photons with ${\bf k}_1$, $\epsilon_1$ and ${\bf k}_2$, $\epsilon_2$ to scatter into two photons with ${\bf k}_3$, $\epsilon_3$ and ${\bf k}_4$, $\epsilon_4$. The zeroth order is given by $\mathcal{M}_0=\delta_{13}\delta_{24}+\delta_{14}\delta_{23}$. The probability for this state to decay into a pair is given by
\be
1-\frac{1}{2}\sum_{3,4}(|\mathcal{M}_0|^2+2\text{Re}\mathcal{M}_0^*\mathcal{M}_4)=-2\text{Re}\mathcal{M}_4 \;,
\ee
where $\mathcal{M}_4$ is illustrated Fig.~\ref{loopsFig}. There is another overall factor of $2$ that comes from the diagrams where the fermion loop goes in the opposite direction. In this approach the field dependence is expressed entirely in terms of the propagator~\eqref{PropagatorInConstantE}, so we do not need the wave functions. Let $s_i$ be the four $s$ integration variables as shown in Fig.~\ref{loopsFig}. The $s_1$ and $s_3$ integrals can be expanded around $s_1=s_3=0$, as in the first approach. We rescale the remaining $s$ variables $s_{2,4}\to s_{2,4}/E$, and then, before performing the ${\bf k}_1$ and ${\bf k}_2$ integrals, we have a saddle point at $s_2=s_4=\text{arccos}\frac{\Sigma_\LCpara}{\sqrt{1+\Sigma_\LCperp^2}}$, where $\Sigma_\mu=(k_1+k_2)_\mu/2$ and $\Sigma_\LCpara=\sqrt{\Sigma_0^2-\Sigma_3^2}$. Let $\delta s_{2,4}$ be the perturbation around these saddle points. With $\delta s_\LCp=(\delta s_2+\delta s_1)/2$ and $\delta s_\LCm=\delta s_2-\delta s_1$ we have
\be
\exp\left\{\frac{1}{E}\left(-\frac{\Sigma}{\sqrt{1-\Sigma^2}}\frac{\delta s_\LCm^2}{2}+2\frac{\sqrt{1-\Sigma^2}}{\Sigma}\delta s_\LCp^2\right)\right\} \;.
\ee
The contour for the $\delta s_\LCp$ integral starts along the real axis up to the saddle point, after which it turns into the imaginary direction. Since only the second half contributes to $\text{Re}\mathcal{M}_4$, we have a factor of $1/2$ compared to a full Gaussian integral (cf.~\cite{Torgrimsson:2018xdf,Callan:1977pt}).

If we in Fig.~\ref{loopsFig} connect the photon line with $-k_i^\mu$ to the one with $k_i^\mu$, $i=1,2$, we find the same diagrams as if one replaces the free photon propagator with a thermal one which is obtained by adding an on-shell part (cf.~e.g.~\cite{Cox:1984vf}). This can help to relate our results to calculations of the effective action.

\subsection{Results}

Collecting everything we find
\be\label{finalP2}
\begin{split}
P_2=&V_4\frac{\alpha^2(\gamma E)^3}{16\pi^2}
\exp\left\{-\frac{2}{E}\left(\frac{\sqrt{\gamma^2-1}}{\gamma^2}+\text{arccsc}\gamma\right)\right\} \\
&\frac{\sqrt{\gamma^2-1}(3\gamma^2-2)+(5\gamma^2-2\gamma^4-2)\text{arccsc}\gamma}{\gamma(\gamma^2-1)^2\text{arccsc}^2\gamma} \;,
\end{split}
\ee
where $V_4$ is a four-dimensional volume factor. 
The exponential part of~\eqref{finalP2} is exactly the same as the one found in~\cite{Brown:2015kgj,Gould:2017fve} without expanding in $\alpha$\footnote{Compare though with the WKB treatment in~\cite{Brown:2015kgj}.}.
The exponential has the form in~\eqref{finalP2} for what~\cite{Gould:2018efv} refers to as intermediate temperatures. As noted in~\cite{Gould:2018efv}, the prefactor in this regime had not been calculated before, so the prefactor in~\eqref{finalP2} is new. In deriving~\eqref{finalP2} we have assumed $\gamma>1$ and, while the exponent has the expected limit as $\gamma\to1$, i.e. $e^{-\pi/E}$, the saddle-point approximation of the prefactor breaks down in that limit. This is not a problem because $P_2$ is anyway small compared to $P_0$ for $\gamma\leq1$, and as far as we are aware there are anyway no results for $P_2$ with $\gamma<1$ that we could have compared with; the two-loop results in~\cite{Gies:1999vb} correspond to $P_1$. 
The prefactor in a different parameter regime has been calculated in~\cite{Gould:2018ovk}, but it has a nontrivial dependence on $\alpha$ and is therefore not something we can directly compare with. 

However, there is a limit in which we can check the prefactor. For $\gamma\gg1$ we expect, e.g. from comparing with similar results for dynamical assistance~\cite{Torgrimsson:2017pzs}, to find a field independent result that agrees with what one finds by setting $E=0$ from the start. This is indeed what we find,
\be\label{P2perturbative}
P_2(\gamma\gg1)=V_4\frac{\alpha^2T^3}{2\pi^2}e^{-\frac{2}{T}} \;,
\ee     
which agrees with Eq.~(8) in~\cite{Gould:2018efv}, see also~\cite{King:2012kd}, for ordinary perturbative Breit-Wheeler pair production summed over photons from a thermal background. On the one hand, it is quite natural that we recover the perturbative result, because $\gamma\gg1$ can be obtained by keeping $T$ fixed while taking $E\to0$, and the exact $P_2$ should of course converge to the perturbative result as the field vanishes. On the other hand, the approximation of the integrals that leads to~\eqref{finalP2} is quite different from the way one would perform the corresponding integrals if $E=0$ from the start, so this agreement is still an interesting and nontrivial check.   

Eq.~\eqref{finalP2} should be compared with the zeroth order, pure Schwinger result
\be\label{Schwinger0}
P_0=V_4\frac{E^2}{4\pi^3}e^{-\frac{\pi}{E}} \;.
\ee 
The exponent in $P_2$ gives an exponential enhancement as soon as $\gamma>1$. However, $P_2$ has a much smaller prefactor because
\be
\frac{\alpha^2E^3}{16\pi^2}\left(\frac{E^2}{4\pi^3}\right)^{-1}\sim4*10^{-5}E<10^{-5} \;,
\ee 
so $\gamma$ has to be sufficiently far above the threshold suggested by the exponent alone, so that the exponential enhancement can overcome the smaller prefactor to give something that is not just on the same order as $P_0$, but something significantly larger. On the other hand, $P_2$ quite quickly converges to its perturbative limit~\eqref{P2perturbative}, so, if one wants something that is significantly different from perturbative Breit-Wheeler, $\gamma$ cannot be too large.
\begin{figure}
\includegraphics[width=\linewidth]{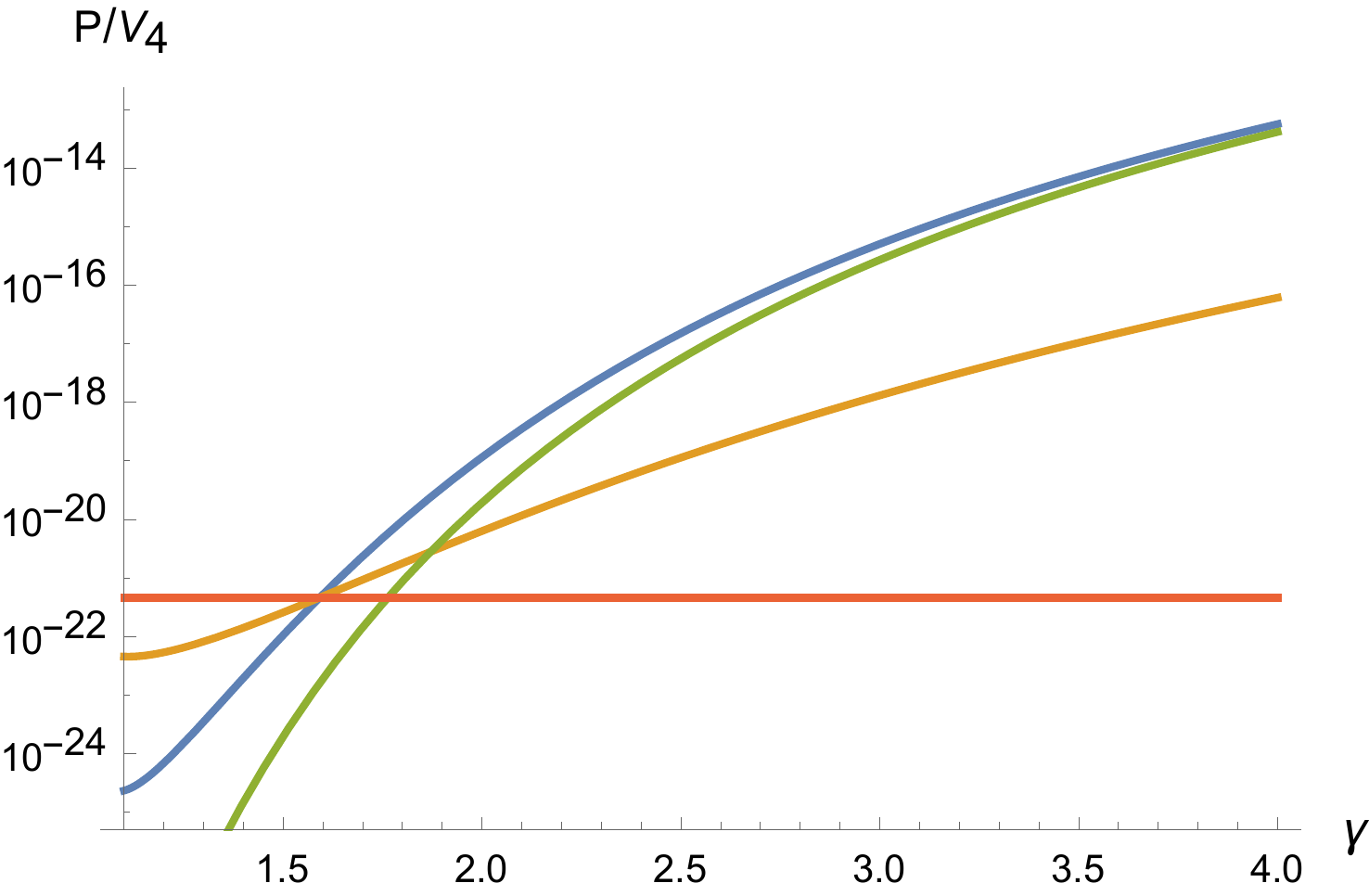}
\caption{Probability at $E=0.08$. The red curve shows $\mathcal{O}(\alpha^0)$~\eqref{Schwinger0}, the orange curve shows $\mathcal{O}(\alpha)$~\eqref{P1p}, the blue curve shows $\mathcal{O}(\alpha^2)$~\eqref{finalP2} and the green curve shows its $\gamma\gg1$ limit~\eqref{P2perturbative}.}
\label{significantDifferenceFig}
\end{figure}
Fig.~\ref{significantDifferenceFig} shows one example of this ``window of significant difference'', where $P_2$ is much larger than $P_0$ as well as its perturbative limit. In this example $P_1$ never gives the dominant contribution, because just above the threshold the exponential enhancement is not enough to compensate for the prefactor suppression compared to $P_0$, and for larger $\gamma$ its exponent grows slower than $P_2$.
These approximations suggest that if we let $E$ be sufficiently large then $P_1$ could become important. However, it is not clear if our approximations are good for such a large $E$. 

Another important point is that our perturbative approach allows us to see that the photons that give the dominant contribution have frequencies on the order of $\omega_{\rm dom}=\sqrt{1-\frac{1}{\gamma^2}}\lesssim1$, i.e. on the order of the electron mass, even though the temperature is low, $T=E\gamma/2\ll1$. So, in this context, the distribution $1/(e^{\omega/T}-1)$ is good if it accurately describes the content of photons with energies on the order of the electron mass. If the distribution instead falls off faster than $e^{-\omega/T}$, then one can expect a significant difference, as shown in~\cite{Torgrimsson:2017pzs,Torgrimsson:2018xdf} for dynamical assistance, where the dominant contribution can come from higher orders.

\section{Higher orders}\label{HigherOrdersSection}

We showed in~\cite{Torgrimsson:2017pzs,Torgrimsson:2018xdf} for dynamical assistance that the dominant contribution can in general come from higher orders, but for a Sauter pulse all higher orders have the same exponential, namely the one in~\eqref{SauterExponent}, which means that the first order gives the dominant contribution. In the case of thermal assistance we have just showed that the second order dominates over the first order for sufficiently weak fields. However, this is because here we are dealing with on-shell photons and the second order is the first order at which the total absorbed momentum can have zero spatial part, and it is the first order which is nonzero even without the electric field. At higher orders we can also have $\sum_{i=1}^N{\bf k}_i=0$ and then the comparison with dynamical assistance suggests that higher orders should have the same exponential as $P_2$. To show this we use the approach in~\cite{Torgrimsson:2018xdf}. The starting point is
\be
\begin{split}
M_n=&(2\pi)^3\delta^3\left({\bf p}+{\bf p'}-\sum_{i=1}^n{\bf k}_i\right)e^n\frak{A}_n \\
&\sim
\int\ud^4x_1...\ud^4x_n
\bar{u}(t_1)e^{ip_jx_1^j}
\slashed{\epsilon}_1e^{-ik_1x_1}G(x_1,x_2) \\ 
&\hspace{1cm}\slashed{\epsilon}_2e^{-ik_2x_2}G(x_2,x_3)\dots \slashed{\epsilon}_{n-1}e^{-ik_{n-1}x_{n-1}} \\
&\hspace{1cm}G(x_{n-1},x_n)\slashed{\epsilon}_n e^{-ik_nx_n}v(t_n)e^{ip'_jx_n^j} \;.
\end{split}
\ee 
For $\sum_{i=1}^N{\bf k}_i=0$ and $p_\LCperp=p'_\LCperp=0$ we can obtain the exponential part by following the same steps as in~\cite{Torgrimsson:2018xdf}: We first perform the ${\bf x}_k$ integrals, which give delta functions, and we change variables $t_k\to(\tau_k+p_3)/E$. 
We expand all the $s_k$ integrals around $s_k\sim0$.
We perform the integral over $\tau_1$ and then the one over $q^{(1)}_1$ (momentum variable for $G(x_1,x_2)$), both with the saddle-point method. Then we perform the integrals over $\tau_2$ and $q^{(2)}_0$, and so on. This gives
\be
|M_n|^2\sim\exp\left\{-\frac{2}{E}\left(\text{arccos}\Sigma-\Sigma\sqrt{1-\Sigma^2}\right)\right\} \;,
\ee 
where 
$\Sigma=\frac{1}{2}\sum_{i=1}^n\omega_n$. The Boltzmann factor also only depends on this sum to leading order,
\be
\prod_{i=1}^n\frac{1}{e^{\omega_n/T}-1}\approx e^{-2\Sigma/T} \;.
\ee 
Compare this with the WKB treatment in~\cite{Brown:2015kgj}.
So, we can estimate the remaining integrals with the saddle point for this sum, $\Sigma_s=\sqrt{1-\frac{1}{\gamma^2}}$, and then we find 
\be
P_n\sim\alpha^n \exp\left\{-\frac{2}{E}\left(\frac{\sqrt{\gamma^2-1}}{\gamma^2}+\text{arccsc}\gamma\right)\right\} \;.
\ee
Thus, all higher orders have the same exponential as $P_2$.
This means that $P_2$ gives the dominant contribution because the higher orders are suppressed by higher powers of $\alpha$. Higher orders could be important if one has a thermal distribution that decays faster than the Boltzmann/exponential scaling, like for example a Gaussian decay. In some sense we are fortunate that the usual thermal distribution has this exponential decay, because it means that we only have to calculate the second order, and the exponential is exactly the same as the one previously obtained with different methods, which gives us a clear check.

\section{Conclusions}

We have studied thermally assisted Schwinger pair production by a Furry picture expansion in $\alpha$. This has allowed us to use the perturbative methods we have developed in previous papers for dynamically assisted Schwinger pair production~\cite{Torgrimsson:2017pzs,Torgrimsson:2017cyb,Torgrimsson:2018xdf}. Apart from the fact that in thermal assistance one has an incoherent sum over photon modes, while in dynamical assistance one has a coherent sum, 
we have found that many aspects are very similar, especially for the case where the weak field in dynamical assistance is a time-dependent Sauter pulse, or some other pulse with exponentially decaying Fourier transform. The reason for this is that the Boltzmann distribution also has an exponential decay. In this context this is a wide distribution with a significant amount of high frequency modes. This means that already the absorption of one (in dynamical assistance) or two (in thermal assistance) photons from the background provides enough energy to give the dominant contribution. This is a good thing from a computational point of view, because it means that we can calculate the pre-exponential factor without considering higher orders. The perturbative approach also shows that the photons that give the dominant contribution has energies on the order of the electron mass, even if the temperature is low. If one instead has a distribution that decays faster than an exponential, then the dominant contribution could come from higher orders.

In this paper we have considered a constant electric background field. We have found that $\mathcal{O}(\alpha^2)$ gives the dominant contribution above a certain threshold in $\gamma$. This threshold is a bit higher than what the exponential part alone would suggest, because the exponential enhancement first has to compensate for the prefactor which is much smaller than the one at $\mathcal{O}(\alpha^0)$. $\mathcal{O}(\alpha^2)$ should of course be larger than $\mathcal{O}(\alpha^0)$ and $\mathcal{O}(\alpha)$ for a sufficiently weak electric field, because the first two orders vanish without the field. The nontrivial conclusion is that $\mathcal{O}(\alpha^2)$ also gives the dominant contribution in a larger region with $\gamma\gtrsim1$. It would be interesting to see how these results generalize to other field shapes, like for example a constant-crossed plane wave~\cite{King:2012kd} or even a pulsed plane wave.
 
Another extension would be to consider initial states with thermal fermions in addition to thermal photons. Then at $\mathcal{O}(\alpha^0)$ one has the effect considered in~\cite{Thermal1980,Gavrilov:2006jb,Kim:2008em,Gavrilov:2007hq,Fukushima:2014sia,Sheng:2018jwf}, which leads to a suppression (for fermions) because of the Pauli principle. At $\mathcal{O}(\alpha^2)$ we would for example have thermal trident pair production, where a thermal fermion interacts with the electromagnetic background field and emits an intermediate photon which subsequently decays into an electron-positron pair.
In this paper we have showed that the photons that give the dominant contribution have energies close to the electron mass, but their energies are still below the electron mass, which suggests that they should be more important than thermal fermions.   
However, the trident process can scale quadratically rather than linearly in the volume (see e.g.~\cite{Baier,Ritus:1972nf,King:2013osa,Dinu:2017uoj,King:2018ibi} for the zero-temperature constant-crossed plane wave case), so it would be interesting to study how large the trident contribution is compared to the $\mathcal{O}(\alpha^2)$ process considered here.

\acknowledgements

G.~T. thanks Holger Gies for inspiring and useful discussions and for reading and commenting on the manuscript, and Oliver Gould for interesting discussions. G.~T. is supported by the Alexander von Humboldt foundation.

\appendix
\section{Starting point}

In this appendix we collect some well-known formulas (for textbooks see e.g.~\cite{LucaQuantumLight,MandlShaw}), which one can use if one wants to derive~\eqref{P1thermalSum} and~\eqref{P2thermalSum} from the incoherent sum over all states weighted by the density matrix. 
In this paper we only consider thermal photons. A complete set for these states is given by
\be
\ket{\{n\}}:=\prod_{i}\frac{(a_i^\dagger)^{n_i}}{\sqrt{n_i!}}\ket{0} \;,
\ee
where $i$ is an index for the momentum and polarization, $n_i$ is the number of particles in the mode $i$, and the mode operators obey
$[a_i,a_j^\dagger]=\delta_{ij}$. The system is put in a spatial volume $V$ with periodic boundary conditions, which means as usual
\be
\sum_i=V\sum_{\rm pol.}\int\frac{\ud^3{\bf k}}{(2\pi)^3} \;.
\ee
The density matrix for the thermal ensemble is given by
\be
\rho(\{n\})=\bra{\{n\}}\hat{\rho}\ket{\{n\}}=\prod_i\frac{e^{-n_i\omega_i/T}}{Z_i} \;,
\ee
where the partition function is given by
\be
Z_i=\frac{1}{1-e^{-\omega_i/T}} \;.
\ee
The photon field is given by
\be
A^\mu(x)=\sum_i\frac{1}{\sqrt{2\omega_iV}}\epsilon_i^\mu a_i e^{-ikx}+{\rm c.c.} \;.
\ee
The pair production probability is given by
\be
\begin{split}
P=&\sum_{\{n\}}\rho(\{n\})\sum_{\{n'\}}\sum_{e^\LCm e^\LCp}|\bra{\{n'\};e^\LCm e^\LCp}S\ket{\{n\}}|^2 \\
=&\sum_{n=0}^\infty P_n \;,
\end{split}
\ee
where $P_n\propto\alpha^n$.
At $\mathcal{O}(\alpha)$ and $\mathcal{O}(\alpha^2)$ this gives~\eqref{P1thermalSum} and~\eqref{P2thermalSum}, respectively.

\end{document}